\documentclass[conference,10pt]{IEEEtran}
\usepackage{balance}

\usepackage{tikz} 
\usepackage{listings}
\usepackage{lscape}
\usepackage{minted}
\usepackage{dirtree}

\usepackage{enumitem}
\usepackage{graphicx}
\usepackage{placeins}
\usepackage{multicol, latexsym}
\usepackage{blindtext}
\usepackage[skip=5pt,font=small]{caption}
\usepackage{longtable}
\usepackage{csquotes}
\usepackage{amsfonts}
\usepackage{amsmath}
\usepackage{amsthm}
\usepackage{amssymb}
\usepackage{algorithm}
\usepackage{tabularx}
\usepackage{capt-of,lipsum} 
\usepackage{balance}
\usepackage{listings}
\usepackage{comment}
\usepackage{diagbox}
\usepackage{pdflscape}
\usepackage{booktabs}
\usepackage{makecell}

\usepackage{newtxtext}
\usepackage{newtxmath}
\usepackage{array}
\usepackage{multirow} 
\usepackage{colortbl}

\usepackage[top=0.70in, bottom=1.0in, left=0.62in, right=0.62in]{geometry}

\usepackage{subcaption}

\usepackage{mathtools, nccmath}

\usepackage{siunitx}
\DeclareSIUnit[number-unit-product = { }] \dBm{dBm}

\usepackage{algpseudocode}
\usepackage{algorithm}

\usepackage{listing-styles}

\algnewcommand\algorithmicforeach{\textbf{for each}}
\algdef{S}[FOR]{ForEach}[1]{\algorithmicforeach\ #1\ \algorithmicdo}

\usepackage[bookmarks=false]{hyperref}

\definecolor{myblue}{rgb}{0.09,0.20,0.34}
\definecolor{mygreen}{rgb}{0,0.6,0}
\definecolor{mygray}{rgb}{0.98,0.98,0.98}
\definecolor{myorange}{rgb}{0.92,0.49,0.34}
\definecolor{mywhite}{rgb}{1.0,1.0,1.0}

\definecolor{NMR}{RGB}{255,255,86}
\definecolor{MRA}{RGB}{255,231,27}
\definecolor{MRD}{RGB}{178,178,178}
\definecolor{MRIP}{RGB}{188,172,0}
\definecolor{MRS}{RGB}{161,207,106}
\definecolor{NA}{RGB}{228,60,52}
\definecolor{AI}{RGB}{255,255,86}
\definecolor{RMR}{RGB}{162,4,21}
\definecolor{RMD}{RGB}{178,178,178}
\definecolor{RMA}{RGB}{255,231,27}
\definecolor{RMIP}{RGB}{188,172,0}
\definecolor{RMS}{RGB}{161,207,106}
\definecolor{LIGHT_GREY}{RGB}{240,240,240}
\definecolor{lightgray}{rgb}{.9,.9,.9}
\definecolor{darkgray}{rgb}{.4,.4,.4}
\definecolor{purple}{rgb}{0.65, 0.12, 0.82}
\definecolor{darkgreen}{RGB}{30, 142, 20}
\definecolor{hilight}{HTML}{C8E5D2} 

\newsavebox{\mybox}

\usepackage{pifont}
\newcommand{\cmark}{\ding{51}}%
\newcommand{\xmark}{\ding{55}}%

\usepackage[scaled]{helvet}
\usepackage[T1]{fontenc}
\usepackage{helvet}

\usepackage[hyphens]{xurl}

\begin{document}




\title{OpenCSI: Self-Calibration Layer for\\ Heterogeneous Mesh Wireless Sensor Networks}


\DeclareRobustCommand*{\IEEEauthorrefmark}[1]{%
  \raisebox{0pt}[0pt][0pt]{\textsuperscript{\footnotesize #1}}%
}

\author{\IEEEauthorblockN{Karim Khamaisi, Simon Sigg, Bruno Rodrigues} 
\IEEEauthorblockA{Sensing Group SG, Institute of Computer Science in Vorarlberg ICV, University of St. Gallen HSG\\Hintere Achmühlerstraße 1c, 6850 Dornbirn, Austria\\
E-mail: \{karim.khamaisi, bruno.rodrigues\}@unisg.ch, simon.sigg@student.unisg.ch}}



\newtheoremstyle{mydef}
{\topsep}{\topsep}%
{}{}%
{\bfseries}{}
{\newline}
{%
  \rule{\linewidth}{0.4pt}\\*%
  \thmname{#1}~\thmnumber{#2}\thmnote{\ -\ #3}.\\*[-1.5ex]%
  \rule{\linewidth}{0.4pt}}%
\theoremstyle{mydef}
\newtheorem{definition}{Definition}
\newtheorem{protocol}{Step}

            


\maketitle

\begin{abstract}
WiFi CSI sensing models trained in one environment usually fail in another because standard per-session normalization bakes chip- and room-specific artifacts into feature space, requiring fresh calibration for every new room or radio. We propose \emph{OpenCSI}, an abstraction layer that hides these artifacts by exposing each mesh link as a single dimensionless Z-score against its own quiet-period temporal standard deviation. The denominator is learned online from a short empty-room bootstrap and reported with a maturity tag, enabling downstream logic to detect drift and abstain when baselines become unreliable.
We evaluate OpenCSI on binary occupancy across three distinct rooms and three ESP32 generations (S3, C3, C6, spanning 802.11n HT20 and 802.11ax HE20), including a same-room chip swap isolating hardware from geometry. A model trained on one deployment holds a single empty-versus-occupied decision threshold zero-shot across nearly all transfer cells, reaching binary F1 up to 0.99 where standard normalization drops to 0.87 or fails outright, with no target-domain data or retraining. The transfer is scoped to binary presence by construction, as distinguishing static from moving motion requires the absolute magnitude that temporal standard deviation removes. We release the source code and dataset to support reproducible cross-environment CSI research.

\end{abstract}



%

\IEEEpeerreviewmaketitle

\pagestyle{plain} 



\section{Introduction} \label{sec:introduction}

\begin{figure}[!t]
    \centering
    \includegraphics[width=\columnwidth]{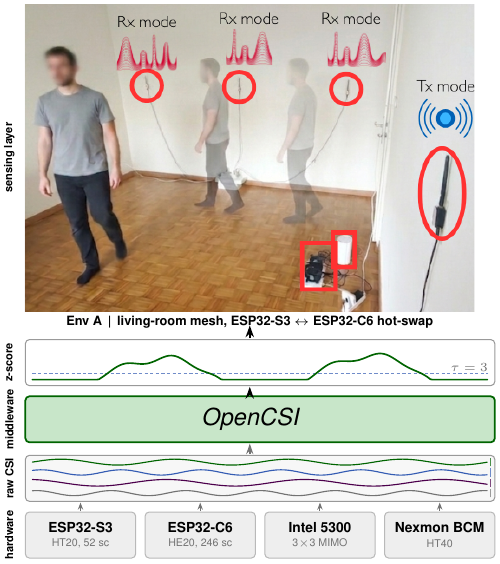}
    \caption{OpenCSI normalises raw CSI from heterogeneous hardware (bottom) into per-link Z-scores (middle). The sensing layer (top, Env~A living-room mesh) consumes one stable input across chip variants.}
    \label{fig:teaser}
\end{figure}

Indoor occupancy sensing has no clean default. Cameras break privacy, Passive Infrared (PIR) misses sitting occupants, and wearables require compliance~\cite{Ahmad2024Occupancy}. Heating, Ventilation, and Air Conditioning (HVAC), lighting, and emergency-response systems still over-ventilate empty rooms and miss occupied ones~\cite{dong2019occupancy}.
WiFi Channel State Information (CSI), the per-subcarrier complex frequency response that an orthogonal frequency-division multiplexing (OFDM) receiver already extracts from every frame, senses motion and presence on hardware already on the wall, with no cameras or wearables~\cite{ma2019wifi,yousefi2017survey}.
Low-cost ESP32 microcontrollers, typically under 3\,USD, expose CSI directly from commodity 802.11 frames~\cite{espressif2023csi}, so a mesh of dozens to hundreds of nodes is now within an undergraduate budget.

Portability is a major challenge for widespread deployment of wireless sensing, because a CSI model trained in one room rarely transfers to another, to a different node batch, or even across a single power cycle of the same node. CSI-Bench~\cite{csibench2025} codifies this barrier, reporting that state-of-the-art methods lose 33-70\% of F1 under cross-device, cross-environment, and cross-user evaluation.
A decade of work has thrown larger models, domain adaptation, and meta-learning at it~\cite{ei2018,crosssense2018,efficientfi2022,dgsense2025,chen2023survey}, each still needing a fresh labeled or unlabeled capture from the new deployment.

The cost in this scenario compounds with scale of the network. A mesh of $N$ nodes exposes $N(N{-}1)$ directed links, where each link carries its own per-boot phase, automatic gain control (AGC) state, and chip-specific gain, and the parameter space multiplies again when chip families are mixed.
The cost comes from standard preprocessing pipelines, which normalize each recording against its own statistics, so the per-boot phase, AGC transient, and chip gain of that recording end up inside the features~\cite{optimalpreproc2023}. A model trained on those features cannot distinguish hardware variations from channel variation, and breaks the moment the hardware also changes.

This paper proposes \emph{OpenCSI}, an abstraction layer that decouples calibration from sensing logic by exposing each link as a single per-link signal (a Z-score).
While the room is empty, each link learns its own quiet-period mean and \emph{temporal} standard deviation through a Welford accumulator~\cite{welford1962}, and once those statistics are warm every frame is converted to a dimensionless Z-score $z = |\bar{A} - \mu|/\sigma$.

Room geometry and chip gain enter $\sigma$ multiplicatively, so the ratio cancels them. The additive noise-floor differences between chip families are not absorbed and compress the ratio in noisier hardware. The deviation $|\bar{A} - \mu|$ in the numerator carries the channel perturbation, so a $2\sigma$ value carries the same meaning across the chips, rooms, and reboots tested, and the sensing layer above consumes one stable per-link signal instead of re-solving calibration on every deployment.

We evaluate \emph{OpenCSI} on binary occupancy across three physically distinct rooms and three ESP32 generations, covering three shifts: cross-room at fixed hardware, cross-chip between rooms, and a controlled cross-generation swap that holds node positions fixed while changing the physical layer (802.11n HT20 to 802.11ax HE20).
Across these deployments a single empty-versus-occupied threshold transfers zero-shot where standard session normalization does not (Fig.~\ref{fig:teaser}, \S\ref{sec:results}).
The abstraction is deliberately bounded: the temporal-std denominator removes absolute magnitude, so the claim covers binary sign-thresholding and stops short of magnitude-dependent tasks like static-versus-moving discrimination (\S\ref{sec:discussion}).

This paper makes three contributions.
\begin{itemize}
    \item \textbf{A single normalized signal per link.} OpenCSI normalizes per-link amplitude by its own quiet-period temporal standard deviation, yielding a chip- and room-independent per-link Z-score. The abstraction holds for binary sign-thresholding across nearly all cross-deployment cells and stops short of magnitude-sensitive tasks; \S\ref{sec:results} quantifies both the transfers and the single failing direction.
    \item \textbf{Calibration drift as a monitorable mesh property.} A per-link reliability score and a frame-level calibration-maturity signal flag stale baselines within seconds, and a short window of fresh quiet observation restores accuracy after drift (\S\ref{sec:drift_recovery}).
    \item \textbf{Open-source.} Apache-2.0 library with the Welford pipeline, maturity tag, and reliability scoring; the three-environment CSI traces (S3, C3, C6 at HT20 and HE20); and the evaluation scripts that reproduce every cell of Tables~\ref{tab:cross_env}--\ref{tab:per_activity}.\footnote{\url{https://github.com/sensing-group/OpenCSI}} \footnote{\url{https://doi.org/10.5281/zenodo.19861032}}
\end{itemize}

\section{Background and Related Work} \label{sec:background}

\begin{figure*}[!t]
    \centering
    \includegraphics[width=\textwidth]{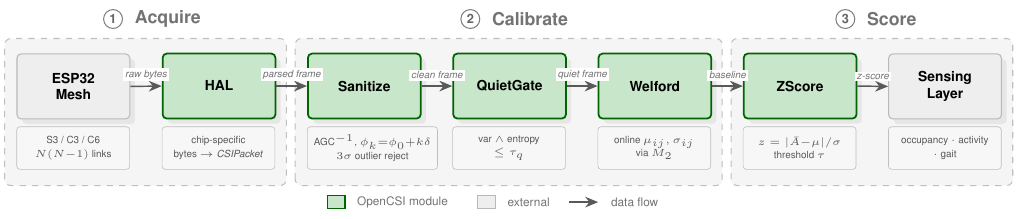}
    \caption{OpenCSI pipeline. Raw CSI from a heterogeneous ESP32 mesh flows through three phase bands (Acquire, Calibrate, Score) and is emitted as the per-link tuple $(z,  \mathrm{reliability})$ to the sensing layer. The dashed arc inside Calibrate marks the QuietGate admitting only quiet-period frames into the Welford estimator.}
    \label{fig:opencsi-pipeline}
\end{figure*}

\begin{table}[!t]
\centering
\caption{Related Work in cross-environment WiFi CSI sensing.}
\label{tab:related_work}
\renewcommand{\arraystretch}{1.18}
\footnotesize
\begin{tabular*}{\columnwidth}{@{\extracolsep{\fill}} l c c c c @{}}
\toprule
 & \multicolumn{2}{c}{\scriptsize\itshape Per-deployment cost} & \multicolumn{2}{c}{\scriptsize\itshape Capability} \\
\cmidrule(lr){2-3}\cmidrule(lr){4-5}
\textbf{Approach} & \makecell{\scriptsize No tgt.\\\scriptsize data} & \makecell{\scriptsize No re-\\\scriptsize training} & \makecell{\scriptsize Cross-\\\scriptsize chip} & \makecell{\scriptsize On-\\\scriptsize device} \\
\midrule
Warmup-normalized amplitude                          & \cmark & \cmark & $\sim^{\dagger}$ & \cmark \\
CSI-ratio~\cite{zengratio2021}                       & \cmark & \cmark & \xmark           & \cmark \\
Domain adapt.~\cite{ei2018,chen2023survey}           & \xmark & \xmark & \xmark           & \xmark \\
Per-deploy.\ calib.~\cite{csibench2025}              & \xmark & \xmark & \xmark           & \cmark \\
Foundation~\cite{scalewhatcounts2025}                & \xmark & $\sim$ & \xmark           & \xmark \\
\midrule
\rowcolor{black!6}
\textbf{OpenCSI}                              & \cmark & \cmark & \cmark$^{\ast}$  & \cmark \\
\bottomrule
\end{tabular*}
\\[2pt]
{\scriptsize \raggedright
$^{\dagger}$ 0.87 F1 cross-environment, but session-specific scale is embedded.\quad
$^{\ast}$ Sign-thresholding; cross-chip confounded with room; C3$\to$S3$^{\prime}$ asymmetric. See \S\ref{sec:results}.\par}
\end{table}

\textbf{Channel State Information} is what an OFDM receiver already produces in the act of decoding a frame. After demodulating the long training field the receiver holds a per-subcarrier complex vector $H_k = A_k e^{j\phi_k}$ that records amplitude and phase across the band, and that richer description is what separates fine-grained sensing from RSSI-style coarse motion~\cite{ma2019wifi,yousefi2017survey,halperin2011tool}. Extracting this vector from commodity hardware once required research-grade network interface cards (NICs) and firmware patches~\cite{halperin2011tool,picoscenes2022,nexmoncsi2019,axcsi2021}, and IEEE~802.11bf will eventually standardize the process~\cite{ieee80211bf}; ESP32 SoCs already expose CSI directly from radio firmware~\cite{espressif2023csi,espidfv55csi} at a price that brings dense meshes within reach.

\textbf{Multipath propagation }(\textit{i.e.,} radio signals travel multiple paths to a receiver, bouncing off objects) turns that vector into a sensor. Frames reach the receiver along a line-of-sight component and many reflected ones, their interference is what $H_k$ encodes, and a moving body rearranges the paths within seconds. The \textbf{Fresnel zones} around each transmitter-receiver pair fix which spatial regions perturb the channel and by how much, so device placement is a first-order design parameter~\cite{fresnel_sensing,Wang2022Placement}. A mesh of $N$ ESP32 nodes exposes $N(N{-}1)$ directed links, each an independent spatial slice of the sensing volume~\cite{espressif2023csi,atif2020esp32}, with 52 subcarriers per link on HT20 and 246 on the ESP32-C6 through the 802.11ax HE-LTF path~\cite{espidfv55csi,axcsi2021}. Adjacent commodity-radio sensing modalities (ZigBee~\cite{wilson2010rti,muller2024big}, BLE~\cite{rodrigues2021bluepil}) share the same passive premise but lack CSI's per-subcarrier resolution.

Calibration is the main challenge that prevents a mesh from transferring across deployments in this scenario. Prior work attacks calibration three ways (Table~\ref{tab:related_work}): algebraic per-frame cancellation, model-side adaptation, and foundation-model pretraining. We compare against each below.

The closest prior method is CSI-ratio~\cite{zengratio2021}, which differs along a complementary axis. CSI-ratio is an \emph{algebraic} per-frame cancellation across the two antennas of a multi-antenna NIC, exact for shared receiver-side gain, carrier frequency offset (CFO), and sampling frequency offset (SFO). OpenCSI is a \emph{statistical} cancellation across each link's own quiet time window, which costs a 120\,s bootstrap but works on single-antenna nodes and across mesh links, where no co-located antenna pair exists.

Most other cross-deployment work pushes the burden into the model. Adversarial domain adaptation~\cite{ei2018}, meta-learning~\cite{crosssense2018}, compressed feature learners~\cite{efficientfi2022}, augmentation-based generalization~\cite{dgsense2025}, body-coordinate-velocity features~\cite{widar3}, and end-to-end deep models~\cite{Turetta2023a} all assume target-environment samples, and the calibration step itself remains per-deployment across the SenseFi~\cite{sensefi2023} and CSI-Bench~\cite{csibench2025} suites. Foundation-model pretraining moves that cost from per-deployment to one-time by learning environment-invariant representations from millions of samples~\cite{scalewhatcounts2025}, but the compute is substantial and cross-hardware generalization remains undemonstrated.


\section{OpenCSI Design} \label{sec:design}

\textit{OpenCSI} is designed to abstract hardware and physical complexities from the sensing layer, sending a single normalized signal per link to the application above (Fig.~\ref{fig:opencsi-pipeline}). This signal is a dimensionless Z-score paired with a scalar reliability tag, aiming to provide a consistent meaning across different hardware, rooms, and reboots.
\textit{OpenCSI} parses hardware-specific packet formats, sanitizes hardware artifacts, and updates per-link baselines during quiet periods.

\subsection{Hardware Abstraction}

Raw CSI packets differ across hardware families in subcarrier count, AGC state encoding, phase noise, and per-boot randomization (S3 HT20: 52 subcarriers in a 240-byte packet; C6 HE20: 246 subcarriers via the IDF~v5.5 HE-LTF path~\cite{espidfv55csi}, following AX-CSI~\cite{axcsi2021}).
A Hardware Abstraction Layer (HAL, as in Fig ~\ref{fig:opencsi-pipeline}) separates parsing from calibration: each hardware family implements a \emph{CSIExtractor} returning a canonical \emph{CSIPacket}, and the downstream pipeline is hardware-agnostic at the API level. We ship and validate extractors for the full ESP32 family (S3, C3, C6); the same interface admits parsers for Intel~5300, Nexmon-patched chipsets, or PicoScenes NICs, but no non-ESP32 traces are reported here, so the empirical claim is scoped to ESP32 silicon.

\subsection{Sanitization}

Sanitization runs three steps before baseline learning. AGC correction inverts the per-packet gain factor recovered from \emph{rx\_state}. A linear phase trend $\phi_k = \phi_0 + k\delta$ is fit and removed. Packets drifting beyond $3\sigma$ from the short-term amplitude median are discarded as retransmits or partial captures.

\subsection{Quiet Period Detection and Baseline Learning}

Calibration learns what the channel looks like at rest by automatically identifying unoccupied periods.
\textbf{Bootstrap (first 120\,s after power-on)} unconditionally treats incoming frames as quiet so the per-link Welford estimators reach stable mean/std; the room \emph{must} be empty during this window or the occupant's perturbation is absorbed into the baseline.
The 120\,s window is a conservative cold start that lets the per-link standard deviation settle before the sensing layer trusts the ratio. Re-anchoring an already-warm accumulator after drift needs only 30\,s (\S\ref{sec:drift_recovery}), because its sample-count and recency terms re-saturate rather than build from zero.
\textbf{Active detection} then gates on two complementary signals over a sliding window: \emph{temporal variance} (low variance = stable channel) and \emph{spectral entropy} of the per-subcarrier amplitude distribution (gates out periodic HVAC/electronic interference that confounds variance-only detectors).
A frame is quiet only if neither signal exceeds its learned threshold; in known-quiet phases (our protocol) the gate can be forced for a specified duration to avoid starving the learner in noisy environments.

When a frame is classified as quiet, each link's sanitized amplitude and phase update a Welford online accumulator~\cite{welford1962} that maintains a running mean and standard deviation without storing raw samples.
Baselines are frequency-bucket-aware (six packet-rate buckets, 2\,Hz to 60+\,Hz) and time-of-day-aware (four 6-hour buckets per day to absorb diurnal thermal drift).

\subsection{Z-score and Reliability: The Per-Link Metric}

Let $(\mu_{ij}, \sigma_{ij})$ denote the per-link mean and standard deviation of the cross-subcarrier amplitude average, learned during quiet periods. The calibrated amplitude Z-score for link $(i \to j)$ at time $t$ is:

\begin{equation}
    z^{\text{amp}}_{ij}(t) = \frac{\left|\bar{A}_{ij}(t) - \mu_{ij}\right|}{\sigma_{ij}},
    \label{eq:zscore}
\end{equation}

where $\bar{A}_{ij}(t)$ is the mean amplitude across subcarriers at time $t$.
The deviation is \emph{unsigned}: a person can increase or decrease received amplitude depending on Fresnel-zone position, so only its magnitude carries presence information.

An analogous phase Z-score is computed from the standard deviation of phase differences across subcarriers, and the primary output \emph{combined\_change} is the signal-to-noise-ratio (SNR) weighted combination
\begin{equation}
    z^{\text{comb}}_{ij} = (1 - w_{\phi}) \cdot z^{\text{amp}}_{ij} + w_{\phi} \cdot z^{\text{phase}}_{ij},
    \label{eq:combined}
\end{equation}
where $w_{\phi} = \sigma((\text{SNR}_{ij} - 15)/5)$ up-weights phase on high-SNR links and down-weights it on low-SNR links.
The differential-phase formulation cancels per-boot phase offsets and timing-offset trends that make absolute ESP32 phase unusable.
The phase channel is exposed end-to-end in the API; on single-antenna ESP32 nodes the SNR weighting attenuates it (amplitude carries most of the signal, \S\ref{sec:results}), but it is retained for multi-antenna platforms where ratio cancellation~\cite{zengratio2021} sharpens the phase statistic.

Each link $(i\to j)$ carries a scalar \emph{reliability} score $r_{ij} \in [0,1]$ that combines four saturating factors:
\begin{equation}
    r_{ij} = r^{(\text{age})}_{ij} \cdot r^{(N)}_{ij} \cdot r^{(\text{SNR})}_{ij} \cdot r^{(\text{rate})}_{ij},
    \label{eq:reliability}
\end{equation}
with $r^{(\text{age})}_{ij} = \exp(-\Delta t_{ij}/\tau_{\text{age}})$ decaying exponentially since the last quiet-period update, $r^{(N)}_{ij} = \min(1, N^{\text{quiet}}_{ij}/N_{\min})$ saturating once the Welford accumulator has $N_{\min}$ quiet samples, $r^{(\text{SNR})}_{ij}$ a sigmoid in instantaneous SNR, and $r^{(\text{rate})}_{ij} = \min(1, f_{ij}/f_{\min})$ saturating at a minimum packet rate. Uninitialized links carry $r_{ij}=0$.
The frame-level \emph{calibration\_maturity} is the weighted mean of $r_{ij}$ across active links; \S\ref{sec:discussion} shows accuracy plateaus once it exceeds ${\sim}0.65$.
The drift-recovery experiment in \S\ref{sec:drift_recovery} bounds the dominant terms empirically: 30\,s of fresh quiet observation restores both $r^{(N)}$ and $r^{(\text{age})}$ to saturation and recovers F1 from 0.25 to 0.98, so under normal operation the deployment threshold is governed by these two factors rather than SNR or packet rate, which change more slowly.

\subsection{Z-score Robustness}
\label{sec:hypothesis}

Eq.~\eqref{eq:zscore} divides each link's per-frame amplitude by its own quiet-period standard deviation.
The carrier frequency, per-subcarrier gain, and hardware AGC all multiply both the numerator and the denominator, so they cancel.
The ratio therefore does not depend on band, Physical Layer (PHY) format, or subcarrier count, and the Z-score should transfer when any of them changes.
We test this prediction in \S\ref{sec:cross-gen} by swapping an 802.11n~HT20 mesh (52 subcarriers) for an 802.11ax~HE20 mesh (246 subcarriers) at identical positions on the same 2.4\,GHz carrier.

Additive noise behaves differently as it enters $\sigma$ as $\sigma^2_{ij} \approx \sigma^2_{\text{channel}} + \sigma^2_n$, and inflates the denominator on hardware with a higher noise floor. The ESP32-C3, for example, has a higher noise floor that compresses the ratio and degrades cross-hardware transfer (\S\ref{sec:results}).

\section{Real-world Evaluation} \label{sec:results}

\begin{figure*}[!t]
    \centering
    \begin{subfigure}[b]{0.32\textwidth}
        \centering
        \includegraphics[viewport=21.6 473.9 153 703.1, clip, height=3.9cm]{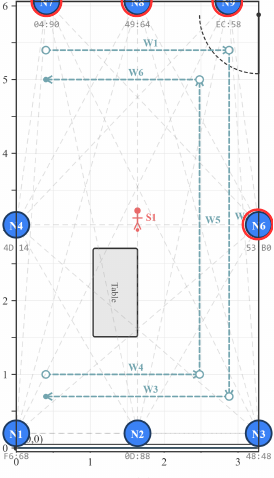}\hspace{2pt}%
        \includegraphics[viewport=167.4 473.9 328 703.1, clip, height=3.9cm]{figures/layout.pdf}
        \caption{Env~A --- 3.28$\times$6.06\,m living room}
        \label{fig:layout-a}
    \end{subfigure}\hfill
    \begin{subfigure}[b]{0.32\textwidth}
        \centering
        \includegraphics[viewport=21.6 229.3 153 438.4, clip, height=3.9cm]{figures/layout.pdf}\hspace{2pt}%
        \includegraphics[viewport=167.4 229.3 328 438.4, clip, height=3.9cm]{figures/layout.pdf}
        \caption{Env~B --- 7.02$\times$4.44\,m robotics lab}
        \label{fig:layout-b}
    \end{subfigure}\hfill
    \begin{subfigure}[b]{0.32\textwidth}
        \centering
        \includegraphics[viewport=21.6 4.8 153 211.5, clip, height=3.9cm]{figures/layout.pdf}\hspace{2pt}%
        \includegraphics[viewport=167.4 4.8 328 211.5, clip, height=3.9cm]{figures/layout.pdf}
        \caption{Env~C --- 2.91$\times$4.43\,m office}
        \label{fig:layout-c}
    \end{subfigure}
    \caption{Floorplans (left of each pair) and photographs (right) of the three deployment environments (hardware variants in Table~\ref{tab:cross_env}). Blue~$\bullet$~=~mesh node; \emph{S}~=~static pose; dashed teal~=~walking path (W1--W6). Photo overlays: mesh nodes, router, and backend host are highlighted in red.}
    \label{fig:layout}
\end{figure*}

\begin{table*}[!t]
\centering
\caption{Binary cross-environment, cross-hardware, and cross-generation transfer (F1-score, RF). Each cell reports F1 for a model trained on the column-labeled deployment and evaluated zero-shot on the block-labeled deployment (train~$\to$~test). Diagonal cells (train~=~test) are the in-session 5-fold reference. Highlighted cells mark the load-bearing zero-shot transfers.}
\label{tab:cross_env}
\resizebox{\textwidth}{!}{%
\begin{tabular}{l|cccc|cccc|cccc|cccc}
\toprule
 & \multicolumn{4}{c|}{\textbf{Test on A (S3)}} & \multicolumn{4}{c|}{\textbf{Test on B (S3$^{\prime}$)}} & \multicolumn{4}{c|}{\textbf{Test on C (C3)}} & \multicolumn{4}{c}{\textbf{Test on A (C6)}} \\
\textbf{Features} & A & B & C & A(C6) & A & B & C & A(C6) & A & B & C & A(C6) & A & B & C & A(C6) \\
\midrule
Calibrated (OpenCSI)    & 1.000 & \cellcolor{hilight}\textbf{0.982} & 0.896 & \cellcolor{hilight}\textbf{0.985} & \cellcolor{hilight}\textbf{0.978} & 1.000 & 0.524 & \cellcolor{hilight}\textbf{0.990} & \cellcolor{hilight}\textbf{0.993} & \cellcolor{hilight}\textbf{0.974} & 1.000 & 0.592 & \cellcolor{hilight}\textbf{0.896} & \cellcolor{hilight}\textbf{0.882} & 0.580 & 1.000 \\
Warmup-normalized       & 1.000 & 0.874 & 0.693 & 0.090 & 0.923 & 1.000 & 0.608 & 0.414 & 0.876 & 0.870 & 1.000 & 0.487 & 0.284 & 0.435 & 0.631 & 1.000 \\
Ablation (cross-sc-std) & 1.000 & 0.848 & 0.655 & 0.105 & 0.917 & 1.000 & 0.690 & 0.451 & 0.876 & 0.789 & 1.000 & 0.437 & 0.623 & 0.429 & 0.446 & 1.000 \\
\bottomrule
\end{tabular}%
}
\end{table*}

\subsection{Setup}

\textbf{Hardware.}
Each environment is instrumented with eight ESP32 nodes in a perimeter topology (corners and mid-walls), yielding 56 directed links.
Env~A and Env~B use \emph{different} ESP32-S3 board batches (802.11n~HT20, 52 subcarriers); we denote the Env~B batch as ESP32-S3$^{\prime}$ to keep the two distinguishable in tables and prose.
All nodes transmit raw CSI via the User Datagram Protocol (UDP) to a central collection host.

\textbf{Environments.}
\begin{itemize}
    \item \textbf{Env~A}: 3.28$\times$6.06\,m living room, training environment, ESP32-S3.
    \item \textbf{Env~B}: 7.02$\times$4.44\,m robotics laboratory in a different building, zero-shot test, ESP32-S3$^{\prime}$.
    \item \textbf{Env~C}: 2.91$\times$4.43\,m office, held-out cross-hardware test, ESP32-C3.
\end{itemize}
Env~C uses ESP32-C3 (same 802.11n~HT20 PHY as S3, different silicon: AGC, noise floor, dynamic range), isolating silicon-level effects at fixed PHY; \S\ref{sec:cross-gen} additionally swaps Env~A's S3 mesh for eight ESP32-C6 nodes (802.11ax~HE20, 2.4\,GHz, 246 useful subcarriers) at identical positions. Floorplans and photographs are in Fig.~\ref{fig:layout}.

\textbf{Data collection.}
Each 12-minute session is recorded at 20\,Hz with phases 5\,min quiet baseline (forced-quiet, encompassing the 120\,s bootstrap of \S\ref{sec:design}) / 3\,min static / 2\,min moving / 2\,min trailing empty, with timestamped class labels (empty, static, moving). Binary evaluation merges static and moving into ``occupied.'' Raw UDP payloads are saved for offline replay.
The quiet segment covers the bootstrap and supplies held-out quiet frames for threshold calibration, the static and moving segments give the occupied sub-states, and the trailing-empty segment tests re-detection of vacancy. Quiet windows were enforced by vacating the instrumented room. Occupancy in adjacent rooms was not controlled, so the reported transfer reflects realistic rather than idealized quiet conditions.

\textbf{Baselines.}
\emph{Warmup-normalized amplitude}: per-link amplitude $A_k = \sqrt{I_k^2 + Q_k^2}$ normalized by the mean/std of the warmup window of the same session, with source statistics applied zero-shot at the cross-environment boundary. We use this as a competence floor for the cross-deployment setting; it is not a claim about the field's strongest preprocessor (e.g., per-subcarrier z-scoring or PCA-on-amplitude), and we revisit that scope in \S\ref{sec:discussion}.
\emph{Ablation (cross-sc-std)}: cross-subcarrier standard deviation only, without OpenCSI's temporal denominator, quiet gating, or reliability weighting; isolates the contribution of the temporal-std denominator.

\textbf{Feature extraction.}
For both OpenCSI and the standard baseline, per-link scalar features are aggregated across the $L$ active links to produce a fixed-size feature vector:
\begin{equation}
\mathbf{f} = \left[\max_l z_l,\ \text{mean}_l z_l,\ \text{std}_l z_l,\ \sum_l \mathbf{1}[z_l > \tau] \right],
\label{eq:features}
\end{equation}
where $\tau = 2.0$ for calibrated Z-scores and $\tau = 1.5$ for normalized amplitude.
OpenCSI additionally provides \emph{reliability}-weighted variants.

\textbf{Models.}
A Random Forest (100 trees) and a two-layer multilayer perceptron (MLP; 64-64, ReLU, Adam) trained in scikit-learn, with 5-fold cross-validation within the source environment for hyperparameters.
We use the Random Forest as the primary learner because its axis-aligned splits are invariant to per-feature rescaling, the property a transferable feature needs, and treat the MLP as a scale-sensitive contrast rather than a competing model (\S\ref{sec:results}).
We report RF throughout; MLP is noted where relevant.

\subsection{Experiment 1: Binary Cross-Environment Transfer}

We train on one environment and evaluate zero-shot on the other with the same model weights.
Table~\ref{tab:cross_env} reports binary F1 (RF) on three transfers: cross-environment at fixed PHY (Env~A~$\leftrightarrow$~Env~B, S3~$\leftrightarrow$~S3$^{\prime}$), cross-hardware at fixed PHY (S3~$\leftrightarrow$~C3 via Env~C), and cross-generation in Env~A (S3~$\leftrightarrow$~C6, \S\ref{sec:cross-gen}).

\begin{figure*}[!t]
    \centering
    \includegraphics[width=\textwidth, trim={40pt 0 0 0}, clip]{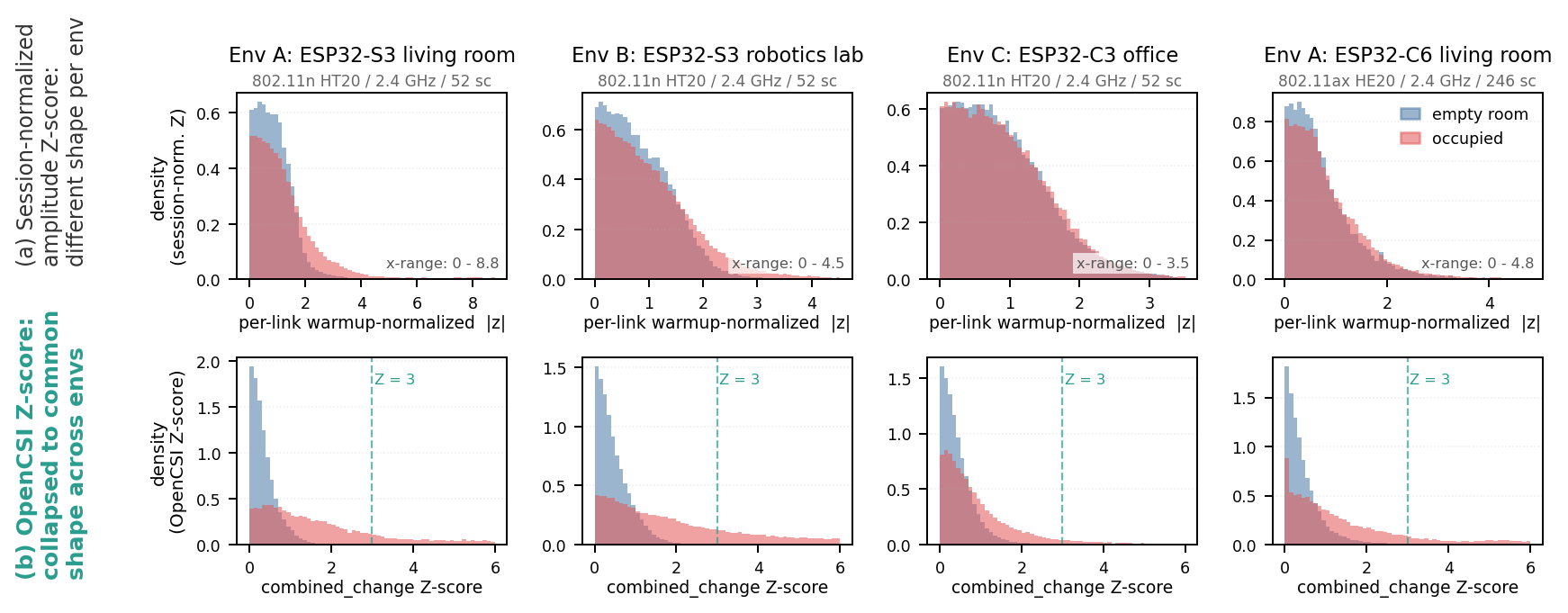}
    \caption{Z-score distributions across deployments (one column per session). Top: standard session-normalized amplitude Z-score, per-deployment range. Bottom: OpenCSI \emph{combined\_change} Z-score, with the empty-vs-occupied split at $z\!\approx\!3$ preserved in every panel. Distributions clipped to the 99.5th percentile.}
    \label{fig:zscore_dist}
\end{figure*}

Calibrated features achieve high cross-environment transfer in the same-variant directions (A$\leftrightarrow$B, S3$\leftrightarrow$S3$^{\prime}$, F1~$\geq$~0.978) and in the cross-variant S3$\to$C3 direction (A,B$\to$C, F1~=~0.993 and 0.974), while warmup-normalized amplitude drops by 8--20 percentage points in the same settings.
The ablation feature set performs no better than warmup-normalized amplitude in several directions, consistent with the temporal-std denominator carrying most of the transfer benefit on the tasks studied.
Figure~\ref{fig:zscore_dist} visualizes the distribution-level effect: standard session-normalized Z-scores have deployment-specific ranges (and in Env~C fail to separate empty from occupied at all), while OpenCSI's calibrated Z-scores retain a single decision threshold ($z\!\approx\!3$) across rooms, hardware variants, and WiFi generations. The cross-generation panel (rightmost, ESP32-C6) is discussed in \S\ref{sec:cross-gen}.

One direction is a structural failure: C(C3)$\to$B(S3$^{\prime}$) drops to 0.524, below both warmup-normalized amplitude (0.608) and the ablation (0.690).
Env~C's C3 nodes produce calibrated Z-scores at a compressed scale ($z_{\max}$~=~7.69 vs.\ 20.47 in A and 30.37 in B), so a classifier trained on the narrower scale does not generalize to the wider one, whereas B$\to$C transfers well. Env~C is the only ESP32-C3 deployment and also the smallest room, so hardware and room axes are confounded; an S3-in-Env-C control is left to future work.

On these single-antenna nodes the SNR weighting in Eq.~\eqref{eq:combined} attenuates the noisier phase channel, so amplitude carries most of the signal (\S\ref{sec:design}).

Classifier choice matters.
A two-layer MLP (64-64) at the same feature dimensionality on the same calibrated features drops to F1~=~0.675 cross-environment (A$\to$B), a 30-point gap behind the Random Forest's 0.978.
The MLP overfits in-domain absolute-scale cues that the Random Forest's piecewise partitioning resists.
The transfer property holds for the Random Forest, not for deep classifiers on the same features. Recovering it on an MLP is a separate question we do not address.

\subsection{Cross-Generation Test: 802.11n HT20 $\to$ 802.11ax HE20}
\label{sec:cross-gen}

Section~\ref{sec:hypothesis} predicts that the dimensionless ratio in Eq.~\eqref{eq:zscore} is invariant to subcarrier count and PHY format.
A clean test requires changing those two quantities without touching the room.
We do so by replacing the eight ESP32-S3 nodes of Env~A with eight ESP32-C6 nodes at identical $(x,y)$ positions: the C6 is a 2.4\,GHz-only WiFi~6 part, so the carrier is unchanged and the only physical-layer differences are 802.11n~HT20$\to$802.11ax~HE20 and 52$\to$246 useful subcarriers.
The same 12-minute protocol is re-run, and the model is reused across the boundary in both directions.

The abstraction holds in both directions, with C6$\to$S3 cleaner than S3$\to$C6.
In the A(C6) row and column of Table~\ref{tab:cross_env}, the C6-trained classifier transfers to S3 at F1~=~0.985 (within 0.024 of the in-session 5-fold reference on A(C6), F1~=~0.961), and the S3-trained classifier transfers to C6 at F1~=~0.896.
Both directions are an order of magnitude above warmup-normalized amplitude (0.09 and 0.28).
The standard top row of Figure~\ref{fig:zscore_dist} fails entirely in both cross-generation directions (F1~$<$~0.30), pointing to the temporal-std denominator as the component that carries the signal across the PHY change rather than any incidental cross-PHY similarity.

The asymmetry between the two directions is residual and visible in the rightmost panel of Figure~\ref{fig:zscore_dist}: C6 empty-room \emph{combined\_change} has a heavier upper tail than S3 ($p_{99.5}$~=~16.4 vs.\ 3.2), so an S3-trained decision boundary occasionally misclassifies C6 quiet frames as occupied; the occupied clusters overlap closely, so the $z\!\approx\!3$ split holds in both directions despite the 4.7$\times$ subcarrier-count change.
We report this as a single-session pilot.

\subsection{Per-Activity Binary Breakdown}

Table~\ref{tab:per_activity} disaggregates binary detection by activity type, revealing where calibration provides the largest gain.
Same-PHY A$\leftrightarrow$B detects both static and moving occupancy reliably with calibrated features (Cal~F1~$\geq$~0.959 on all six cells); warmup-normalized amplitude degrades most on empty-vs-moving (0.869 vs.\ 0.964 in A$\to$B), consistent with movement-induced variance being more environment-sensitive.

The cross-generation direction C6$\to$\{A,B\} is uniformly strong (Cal~F1~$\geq$~0.964 on all three tasks); the reverse \{A,B\}$\to$C6 is asymmetric, with the largest gap on empty-vs-moving (Cal~A$\to$C6~=~0.611, B$\to$C6~=~0.806), the residual S3$\to$C6 scale shift already discussed in \S\ref{sec:cross-gen}.
On S3$\to$C3, calibrated B$\to$C empty-vs-moving F1 falls to 0.467 (vs.\ 0.593 standard) under the same C3-narrow-range mechanism; A$\to$C and the reverse directions transfer cleanly.

\begin{table*}[!ht]
\centering
\caption{Per-activity binary F1-score (RF) along three transfer axes: cross-environment same-PHY (A$\leftrightarrow$B), cross-generation in the same room (A$\leftrightarrow$A(C6)), and joint cross-environment + cross-generation (B$\leftrightarrow$A(C6)). Cal~=~calibrated OpenCSI features, Std~=~warmup-normalized amplitude; each column is a single train~$\to$~test direction. Bold cells mark the strongest calibrated cross-generation entries.}
\label{tab:per_activity}
\resizebox{\textwidth}{!}{%
\begin{tabular}{l|cccc|cccc|cccc}
\toprule
 & \multicolumn{4}{c|}{\textbf{A $\leftrightarrow$ B (cross-env, S3 $\leftrightarrow$ S3$^{\prime}$)}} & \multicolumn{4}{c|}{\textbf{A $\leftrightarrow$ A(C6) (cross-generation)}} & \multicolumn{4}{c}{\textbf{B $\leftrightarrow$ A(C6) (cross-env + cross-gen)}} \\
\textbf{Task} & \textbf{Cal A$\to$B} & \textbf{Cal B$\to$A} & \textbf{Std A$\to$B} & \textbf{Std B$\to$A} & \textbf{Cal A$\to$C6} & \textbf{Cal C6$\to$A} & \textbf{Std A$\to$C6} & \textbf{Std C6$\to$A} & \textbf{Cal B$\to$C6} & \textbf{Cal C6$\to$B} & \textbf{Std B$\to$C6} & \textbf{Std C6$\to$B} \\
\midrule
Empty vs.\ static       & \cellcolor{hilight}0.986 & \cellcolor{hilight}0.987 & 0.900 & 0.956 & 0.717 & \cellcolor{hilight}\textbf{0.968} & 0.512 & 0.409 & 0.935 & \cellcolor{hilight}\textbf{0.993} & 0.421 & 0.412 \\
Empty vs.\ moving       & \cellcolor{hilight}0.964 & \cellcolor{hilight}0.959 & 0.869 & 0.880 & 0.611 & \cellcolor{hilight}\textbf{0.964} & 0.659 & 0.572 & 0.806 & \cellcolor{hilight}\textbf{0.979} & 0.668 & 0.482 \\
Empty vs.\ all occupied & \cellcolor{hilight}0.982 & \cellcolor{hilight}0.978 & 0.874 & 0.923 & 0.703 & \cellcolor{hilight}\textbf{0.980} & 0.575 & 0.359 & 0.880 & \cellcolor{hilight}\textbf{0.990} & 0.457 & 0.370 \\
\bottomrule
\end{tabular}%
}
\end{table*}

The reach of the abstraction stops at sign-thresholding: three-class (empty/static/moving) macro F1 falls below the binary threshold cross-environment because that task requires the absolute magnitude of the deviation, which the denominator in Eq.~\eqref{eq:zscore} absorbs by design (\S\ref{sec:discussion}, sign-vs-magnitude).

\subsection{Calibration Drift: Detection and Online Recovery}
\label{sec:drift_recovery}

A deployed mesh's calibration can go stale (a node restart, furniture motion, or a baseline migrated from a structurally similar room). A baseline-swap ablation on A$\to$B replacing Env~B's per-link baseline with a pooled Env~A one collapses binary F1 from 0.98 to 0.25 with empty-class recall at 0.5\%: the baseline mean offset swamps any real perturbation, so every B frame looks uniformly active.

The collapse is detectable on-device: \emph{calibration\_maturity} (\S\ref{sec:design}) drops below the $\sim$0.65 operational threshold (\S\ref{sec:discussion}) within seconds, so the mesh rejects on failure rather than silently producing wrong outputs. Recovery is fast: rebuilding the per-link baseline from 30\,s of fresh quiet observation restores F1 to 0.98 (15\,s already gives 0.976).

Per-deployment calibration is required but recoverable in 30\,s, so staleness is a transient fault under \emph{calibration\_maturity} monitoring. Because each of the 56 directed links carries its own \emph{reliability} score (\S\ref{sec:design}), and \emph{calibration\_maturity} is their weighted mean, a drop also pinpoints \emph{which} links have degraded.

\section{Deployment Analysis} \label{sec:discussion}

\subsection{Calibration Maturity and Warmup Time}

\begin{figure}[!t]
    \centering
    \includegraphics[width=\columnwidth]{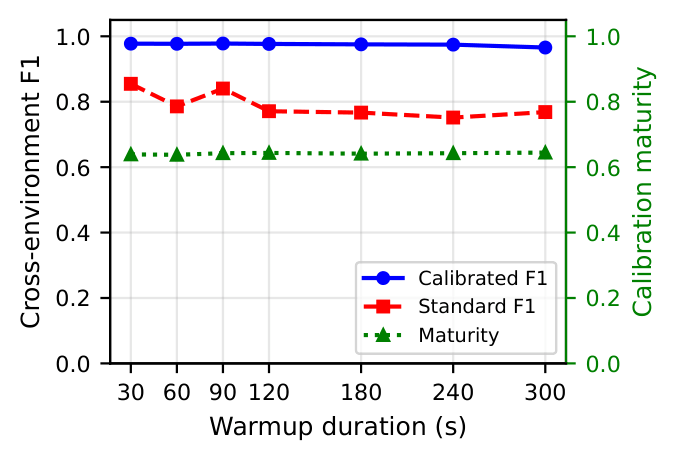}
    \caption{Cross-environment F1 and OpenCSI calibration maturity vs.\ quiet warmup duration. The 0.65 maturity plateau is empirical to our three environments.}
    \label{fig:maturity}
\end{figure}

How long must a node observe before its Z-scores are usable?
Fig.~\ref{fig:maturity} plots cross-environment F1 (A$\to$B) against warmup duration alongside the frame-level \emph{calibration\_maturity} score.

Maturity reaches $\sim$0.65 after 5 minutes (A: 0.67, B: 0.65, C: 0.62), once the per-link Welford accumulators are stable. Calibration state persists to disk, so warmup is paid once per deployment, not per restart.

\subsection{Sign vs.\ Magnitude}

The temporal-std denominator in Eq.~\eqref{eq:zscore} absorbs the gain that a particular hardware and geometry apply to channel fluctuations, so what crosses any fixed threshold on the ratio is whether a perturbation occurred rather than how large it was in absolute units. Empty-vs-occupied detection therefore reduces to a sign-style threshold and transfers across hardware and rooms, whereas static-vs-moving discrimination depends on residual magnitudes that still scale with the hardware noise floor and room geometry and does not.

The empirical pattern matches the prediction: empty $z_{\max}$ holds at 1.17, 1.85, and 1.90 across environments A, B, and C, well below any occupied state, while static and moving $z_{\max}$ scale with room and hardware by up to a factor of four yet preserve the static~$>$~moving ordering within each deployment. 

\subsection{Deployment and Scaling}

Bringing OpenCSI online takes three steps: power on the mesh in an empty room, wait until maturity climbs above 0.65 (about 5 minutes), and let the library checkpoint to disk. The sensing layer then reads $(z, r)$ tuples per link. A maturity drop flags drift (a node restart, a moved baseline, rearranged furniture), and 30\,s of fresh quiet observation rebuilds the affected statistics (\S\ref{sec:drift_recovery}).

Quiet periods are the resource the method consumes. Office and residential deployments provide them at night and during lunch hours; the time-of-day buckets absorb diurnal thermal drift. Server rooms and factories lack natural quiet windows and would need opportunistic low-motion gating or a pre-loaded baseline. Slow large-object motion (an open door, moved furniture) sits between the variance and spectral-entropy gates and can be absorbed into the baseline; the maturity drop flags it but does not recover it.

Compute scales with link count, not subcarrier count. Per-frame work is $O(K)$ to average across $K$ subcarriers plus a constant per-link update, so per-node cost grows as $O(K(N{-}1))$ at mesh size $N$. Memory is a few kilobytes at $N{=}8$ and stays under 1\,MB at $N{=}50$. Per-link statistics update locally; the bottleneck at scale is mesh airtime, not calibration, and how to allocate that airtime across links has its own dilution-vs-coverage tradeoff that we examine separately~\cite{khamaisi2026dilution}.

The per-link signal does not depend strongly on hardware-specific physics. Weights swap across the S3$\leftrightarrow$C6 hot-swap at identical positions, with C6$\to$S3 transferring at F1~=~0.985 and the reverse at 0.896 (\S\ref{sec:cross-gen}). Because drift is on-device-detectable, deployments can declare a maturity threshold below which the sensing layer abstains rather than fail silently.

\subsection{Limitations}

Our evaluation covers binary presence and three-class activity on ESP32 nodes at 2.4\,GHz, using tree-based learners over amplitude-dominated single-antenna features. Richer outputs like multi-person counting and localization remain open, and the HAL admits parsers for Intel 5300, Nexmon, and PicoScenes that we have not yet exercised in this work. 

5/6\,GHz operation and multi-antenna platforms are next. On the latter, the differential phase becomes useful as antenna ratios cancel its noise sources~\cite{zengratio2021}. The 120\,s bootstrap assumes a known-empty room, and the calibration-maturity signal reports when that assumption is violated, with longer-horizon characterization of multi-day stability and run-to-run variance across source-target pairings left to longer-trace deployments.

\section{Conclusion} \label{sec:conclusion}

OpenCSI is an abstraction layer between hardware and sensing logic: each link is exposed as a single dimensionless Z-score (amplitude divided by its own quiet-period temporal standard deviation), so the sensing layer above need not model chip- or room-specific physics. The abstraction is approximate. The temporal-std denominator divides out chip gain but not noise-floor differences, which compresses the ratio on noisier silicon (\S\ref{sec:results}). 
Future work fills the experimental grid, runs non-ESP32 silicon, and bounds multi-day stability.





\section*{Acknowledgments}
AI tools were used for spelling, grammar, style, and code assistance in prototype development. The conceptualization, methodological design, and interpretation of results originated from the authors.


\bibliographystyle{IEEETranS}
\let\oldthebibliography\thebibliography
\renewcommand{\thebibliography}[1]{%
  \oldthebibliography{#1}%
  \setlength{\itemsep}{-0.1ex}%
}
\bibliography{bib/main.bib}

\end{document}